\documentclass[a4paper,fleqn,usenatbib]{mnras}
\usepackage{newtxtext,newtxmath}

\usepackage[T1]{fontenc}
\usepackage{ae,aecompl}
\usepackage{color}

\usepackage{graphicx}   
\usepackage{amsmath}    
\usepackage{amssymb}    
\usepackage{subcaption}
\title[A New GW Signature of Low-$T/|W|$ Instability]{A New Gravitational Wave Signature of Low-$T/|W|$ Instability in Rapidly Rotating Stellar Core Collapse}

\author[Shibagaki, Kuroda, Kotake, \& Takiwaki]{
Shota Shibagaki,$^{1}$
Takami Kuroda$,^{2}$
Kei Kotake,$^{1,3}$ and 
Tomoya Takiwaki$^{4}$
\\
$^1$Department of Applied Physics, Fukuoka University, 8-19-1, Nanakuma, Fukuoka, 814-0180, Japan\\
$^2$Institut f{\"u}r Kernphysik, Technische Universit{\"a}t Darmstadt, Schlossgartenstrasse 9, D-64289 Darmstadt, Germany\\
$^3$Research Insitute of Stellar Explosive Phenomena (REISEP), Fukuoka University, Nanakuma 8-19-1, Johnan, Fukuoka 814-0180, Japan\\
$^4$Division of Science, National Astronomical Observatory of Japan (NAOJ), 2-21-1, Osawa, Mitaka, Tokyo, 181-8588, Japan
}

\date{Accepted XXX. Received YYY; in original form ZZZ}

\pubyear{2020}

\begin{document}
\label{firstpage}
\pagerange{\pageref{firstpage}--\pageref{lastpage}}
\maketitle

\begin{abstract}
We present results from a full general relativistic three-dimensional hydrodynamics simulation of rapidly rotating core-collapse of a 70 M$_{\odot}$ star with three-flavor spectral neutrino transport.
We find a strong gravitational wave (GW) emission that originates from the growth of the one- and two-armed spiral waves extending from the nascent proto-neutron star (PNS). The GW spectrogram shows several unique features that are produced by the non-axisymmetric instabilities. After bounce, the spectrogram first shows a transient quasi-periodic time modulation at $\sim$ 450 Hz. In the second active phase, it again shows the quasi-periodic  modulation but with the peak frequency increasing with time, which continues until the final simulation time. From our detailed analysis, such features can be well explained by a combination of the so-called low-$T/|W|$ instability and the PNS core contraction.
\end{abstract}

\begin{keywords}
supernovae: general ---  stars: neutron --- hydrodynamics --- gravitational waves
\end{keywords}


\section{Introduction}

The LIGO and Virgo collaborations have opened a new window on the gravitational wave (GW) astronomy. The detailed analyses of the GW signal from compact binary coalescences have constrained not only the binary parameters but also physical quantities related to the neutron star (NS) physics such as the radii and nuclear equation of state (EOS) \citep{Abbott19catalog}. 

Some of the binary black holes (BHs) detected by GWs have their masses of $\sim$30-50 M$_{\odot}$, indicating their formation in low metalicity environments \citep[e.g.,][]{abbott16}. 
Recently, an astronomical observation of \citet{Liu19} reported the detection of a $\sim$70M$_{\odot}$ BH in a binary system (``LB-1'') with its companion being an $\sim$8 $M_{\odot}$ solar metalicity B star. 
These findings motivate not only theoretical studies on the formation channels of these massive BHs 
\citep[e.g.,][see, however, \citealt{Abdul19,Eldridge19,El19}]
{Belczynski19}
, but also numerical simulation studies clarifying the  hydrodynamics processes of the massive BH formation  \citep[e.g.,][]{Chan&Muller18,KurodaT18}.

Next to the compact binary coalescence, core collapse (CC) of massive stars and the subsequent explosions have been considered one of the most promising GW sources. Extensive studies have revealed that the candidate ingredients of GW emission from stellar CC include prompt convection, neutrino-driven convection, proto-neutron star (PNS) convection, $g$/$f$-mode oscillation of the PNS, the standing accretion shock instability, rotational flattening of the bouncing core, and non-axisymmetric rotational instabilities \citep[for reviews, see][]{fryer11,kotake_kuroda16}. Among these multiple physical elements, progenitor rotation has
 been long considered the primary ingredient that leads to the sizable GWs \citep{ewald82}.

In three-dimensional (3D) simulations of rapidly rotating CC, the growth of non-axisymmetric rotational instabilities  that develop later in the postbounce phase has been reported \citep{Ott05,Ott07_prl,Scheidegger07,Scheidegger10,KurodaT14,takiwaki16,takiwaki18}. 
These studies have shown that the so-called low-$T/|W|$ instability \citep{Watts05,saijo06}, where $T/|W|$ is the ratio of the rotational to gravitational potential energy, 
 plays a key role for the growth of the non-axisymmetric flows that extend outward in the vicinity of the PNS. Since the stability criterion depends on the compactness of the PNS, it is important to include general relativity (GR) and self-consistent neutrino transport in 3D rapidly rotating CC simulations. 
This was already pointed out by \citet{Ott07_prl,Ott12,KurodaT14} with different levels of sophistication for  neutrino treatment.  
 So far 3D simulations of rapidly rotating collapse that followed a long-term postbounce evolution with spectral neutrino transport have been reported in  \citet{takiwaki16}
 and \citet{Obergaulinger19}, but with the assumption of Newtonian and post-Newtonian gravity, respectively.
 
In this {\it Letter}, we present a first 3D full-GR hydrodynamics simulation of rapidly rotating CC of a 70 M$_{\odot}$ star
 with spectral neutrino transport, where we follow the longest postbounce evolution 
 ($\sim 270$ ms after bounce) in this context.
 Our results confirm that the growth of the non-axisymmetric instability  leads to long-lasting quasi-periodic GW emission.
As a new intriguing result, the characteristic GW frequency increases with time.
Such a ramp-up feature of the GW frequency may look similar to that from the $g/f$-mode oscillation of the {\it non-rotating} PNS \citep{Murphy09,BMuller13,Vartanyan19}. However, 
in rapidly rotating CC, we show that the ramp-up feature (with much bigger GW amplitudes) is 
produced predominantly because of the growth of the low-$T/|W|$ instability and the PNS contraction.
\section{Numerical Method and Setup}
The numerical schemes of our 3D-GR simulation are essentially the
same as those in \citet{KurodaT16}, except the update where we adopt the directionally unsplit predictor-corrector scheme instead of the use of Strang splitting. 
Based on the BSSN formalism, 
we solve the metric evolution in fourth-order accuracy 
 by a finite-difference scheme in space and with a Runge-Kutta method in time
.
We consider three-flavor of neutrinos ($\nu \in \nu_e,\bar\nu_e,\nu_x$) with 
$\nu_x$ denoting heavy-lepton neutrinos. 
Employing an M1 analytical closure scheme \citep{Shibata11},
 we solve spectral neutrino transport of the radiation energy and momentum including all the gravitational red- and Doppler-shift terms
  using 12 energy bins from 1 to 300 MeV.
 Regarding neutrino opacities, the 
standard weak interaction set in \citet{Bruenn85} plus nucleon-nucleon Bremsstrahlung is taken into account
 (see \citet{KurodaT16} for more detail).

We use a 70 M$_{\odot}$ zero-metallicity star of \cite{Takahashi14}.
At the precollapse phase, the mass of the central iron core is $\sim4.6$ M$_{\odot}$ and the enclosed mass up to the helium layer is $\sim 34$ M$_{\odot}$.
The central density profile of this progenitor is similar to that of 75 M$_\odot$ ultra metal-poor progenitor model of \citet{WHW02} that is used as a collapsar model in \citet{Ott11}.
The central angular frequency of the original progenitor model is $\sim$ 0.03 rad s$^{-1}$, by which one can roughly estimate the $T/|W|$ at bounce as $\sim10^{-5}$  due to the angular momentum conservation.
For such slow rotation, it is very unlikely to produce   non-axisymmetric rotational instabilities, and the GW signature would not be significantly deviated from those from a non-rotating model.
Therefore, to explore the impact of rotation, we assume a rapidly rotating iron core with the central angular velocity being 2 rad s$^{-1}$ \citep[e.g., as taken in ][]{Scheidegger08,Ott11}.
For reference, the initial rotational energy and $T/|W|$ are $\sim1.3\times10^{45}$ ergs and $\sim2\times10^{-7}$ for the original progenitor model and are $\sim2.6\times10^{49}$ ergs and $\sim3\times10^{-3}$ for the rapidly rotating model explored in this work, respectively.

We use the EOS by \citet{LSEOS} with a bulk incompressibility
modulus of $K$ = 220 MeV (LS220). 
The 3D computational domain is a cubic box
with 15,000 km width, and nested boxes with nine refinement
levels are embedded in the Cartesian coordinates. Each box contains $64^3$ cells and the
minimum grid size near the origin is $\Delta x=458$m.
The PNS core surface ($\sim10$ km) and stalled shock ($\sim110$-300km)are resolved by $\Delta x=458$m and $7.3$ km, respectively.
In the numerical analysis below, we extract GWs with a standard quadrupole formula including GR corrections \citep{Shibata&Sekiguchi03,KurodaT14}. The
GW spectrograms are obtained using short-term Fourier transformation with a Hann window, whose width is set as 20 ms. $t_{\rm pb}$ represents the time measured after core bounce.

\section{Result}
\begin{figure*}
\includegraphics[bb= 0 0 470 360, width=0.49\textwidth]{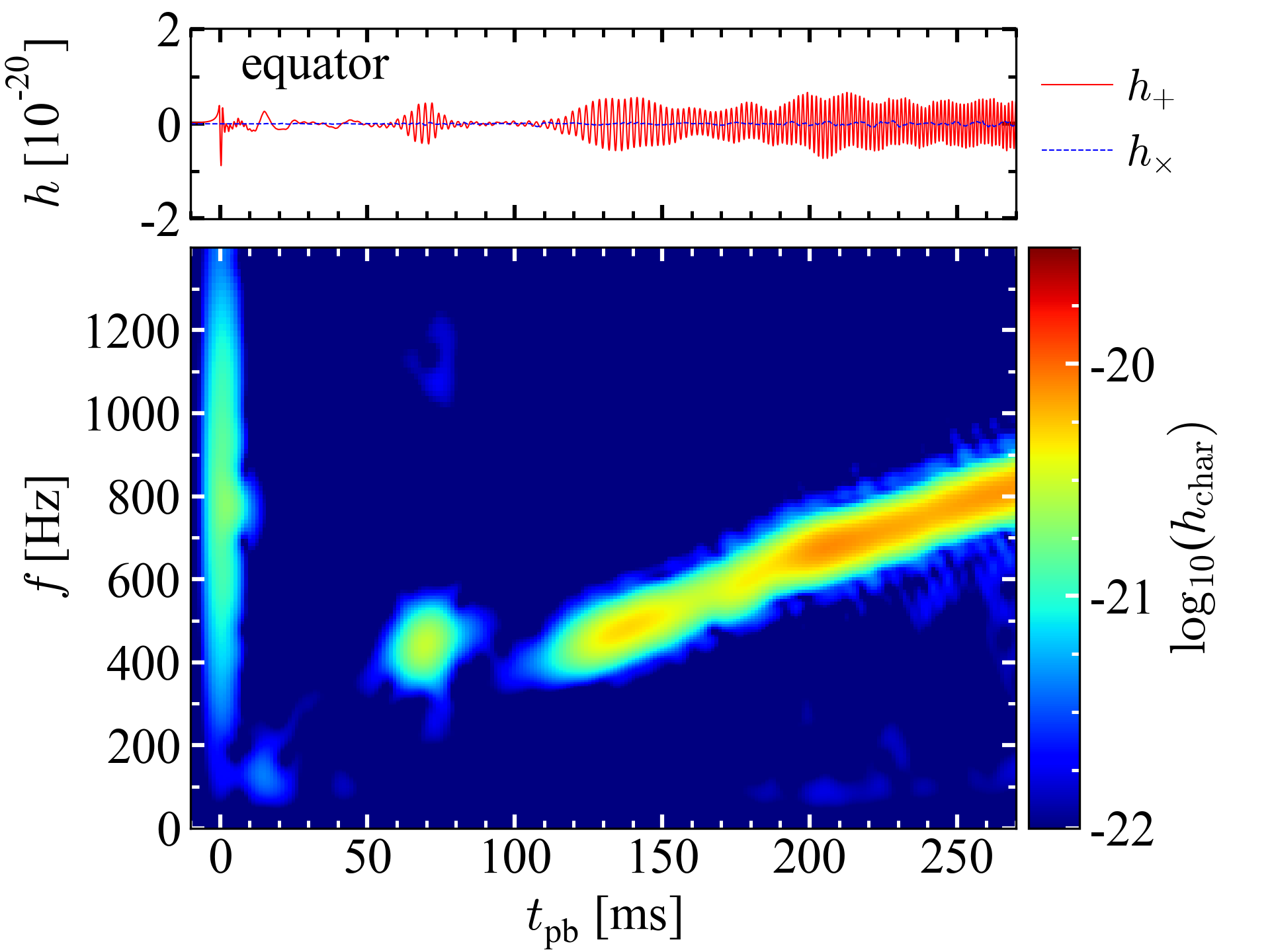}
\includegraphics[bb= 0 0 470 360, width=0.49\textwidth]{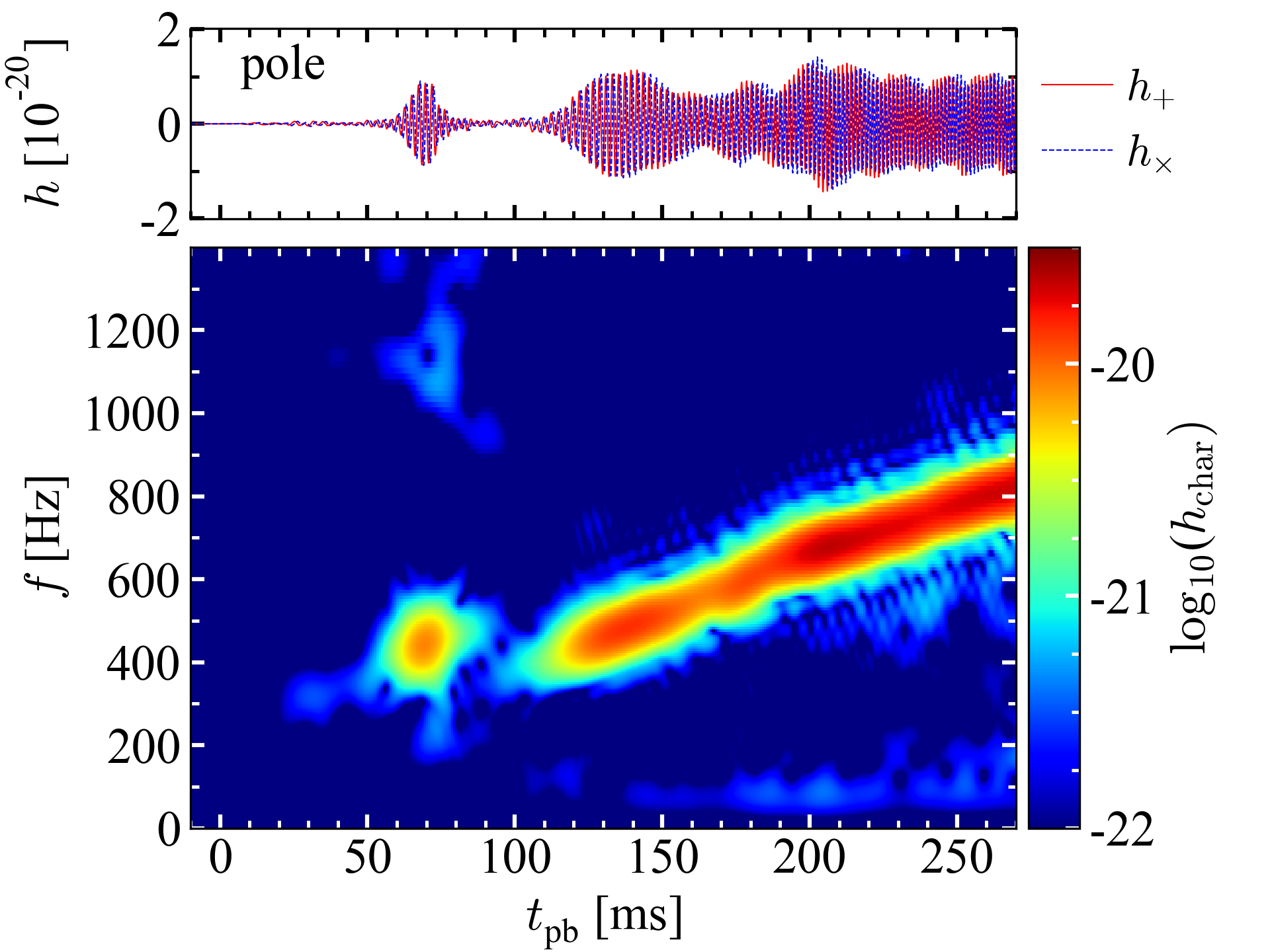}
\caption{GW strains (top) and spectrograms of their characteristic strains (bottom) seen along the equator (left) and along the pole (right) at a source distance of 10 kpc. The plus modes and the cross modes of the GW strains are shown by the red solid lines and the blue dashed lines, respectively.\label{fig:strain}}
\end{figure*}

We begin with a brief description of the post-bounce dynamical evolution in our simulation.
The central density reaches $\sim 3.6\times 10^{14}$ g cm$^{-3}$ at core bounce.
After core bounce and the subsequent ring-down oscillation, the central density monotonically increases up to $\sim$5$\times 10^{14}$ g cm$^{-3}$ at the end of the simulation ($t_{\rm{pb}} =$ 270 ms). 
The bounce shock propagates outward to reach $\sim$ 200 km at $t_{\rm{pb}} \sim$ 70 ms.
As we will see later, the low-$T/|W|$ instability takes place at this time and temporally accelerates the shock.
But once the instability ceases, the shock propagation decelerates and stagnates at $\sim$ 300 km at $t_{\rm{pb}} \sim$ 85 ms.
Afterward the shock surface starts shrinking.
The shrink stops at $t_{\rm{pb}} \sim$ 100 ms, and later the average shock radius keeps $\sim$ 200 km until the end of our simulation.

 Fig. \ref{fig:strain} shows the GW strain $h$ (top panels) and the spectrograms
 of the characteristic strain ($h_{\rm{char}}$, bottom panels, see the definition of equation (44) in \citealt{KurodaT14}) 
for a source distance of $D=$10 kpc. For convenience, we denote $h_{+}$ (red solid line) and $h_{\times}$ (blue dashed line) with subscripts $I$ and $II$, corresponding to the (reference) viewing direction either along the rotational axis (the positive $z$-axis, right panels) or equatorial plane  (the positive $x$-axis, left panels), respectively.
After core bounce and the subsequent ring-down phase, the waveforms 
show a quasi-periodic time modulation around $t_{\rm pb} \sim 70 $ ms. 
This can be seen as a clear excess in the spectrograms
with the peak frequency of $f \sim 450$ Hz. 
The GW amplitude is more strongly emitted toward the pole than the equator. 
The waveforms show that this lasts only for the duration of $\sim 20$ ms followed by a quiescent phase until $t_{\rm{pb}} \sim 110$ ms, after which the quasi-periodic GW emission becomes active again.

In the second active phase, the GW emission does not subside quickly as seen in the first phase. The spectrograms clearly show that the peak GW frequency, either seen from the equator or pole, increases with time from $f \sim 450$ Hz at $t_{\rm pb} \sim 120$ ms to $\sim 800$ Hz at the end of our simulation time ($t_{\rm pb} \sim 270$ ms).
Seen from the pole, the phase difference between $h_+$ and $h_{\times}$ is $\pi/2$. 
Furthermore, the GW amplitudes seen from the pole are approximately two times bigger than those from the equator, i.e.,  $h_{+/\times,I}\sim 2 h_{+,II}$, and the cross mode of the GWs seen from the equator ($h_{\times,II}$) is significantly smaller than the plus mode ($h_{+,II}$) . 
These GW features are consistent with previous 3D studies where the growth of non-axisymmetric instabilities were observed \citep{Ott05,Scheidegger07,Scheidegger10,KurodaT14,takiwaki18}.
We note that the GW amplitudes are bigger than those from a 75 M $_\odot$ model in
 \citet{Ott11} with the same initial angular momentum, where the growth of the low-$T/|W|$ instability was not fully captured by enforcing the octant symmetry in their simulation.

\begin{figure}
\centering
\includegraphics[bb= 720 200 2220 1450, width=0.235\textwidth]{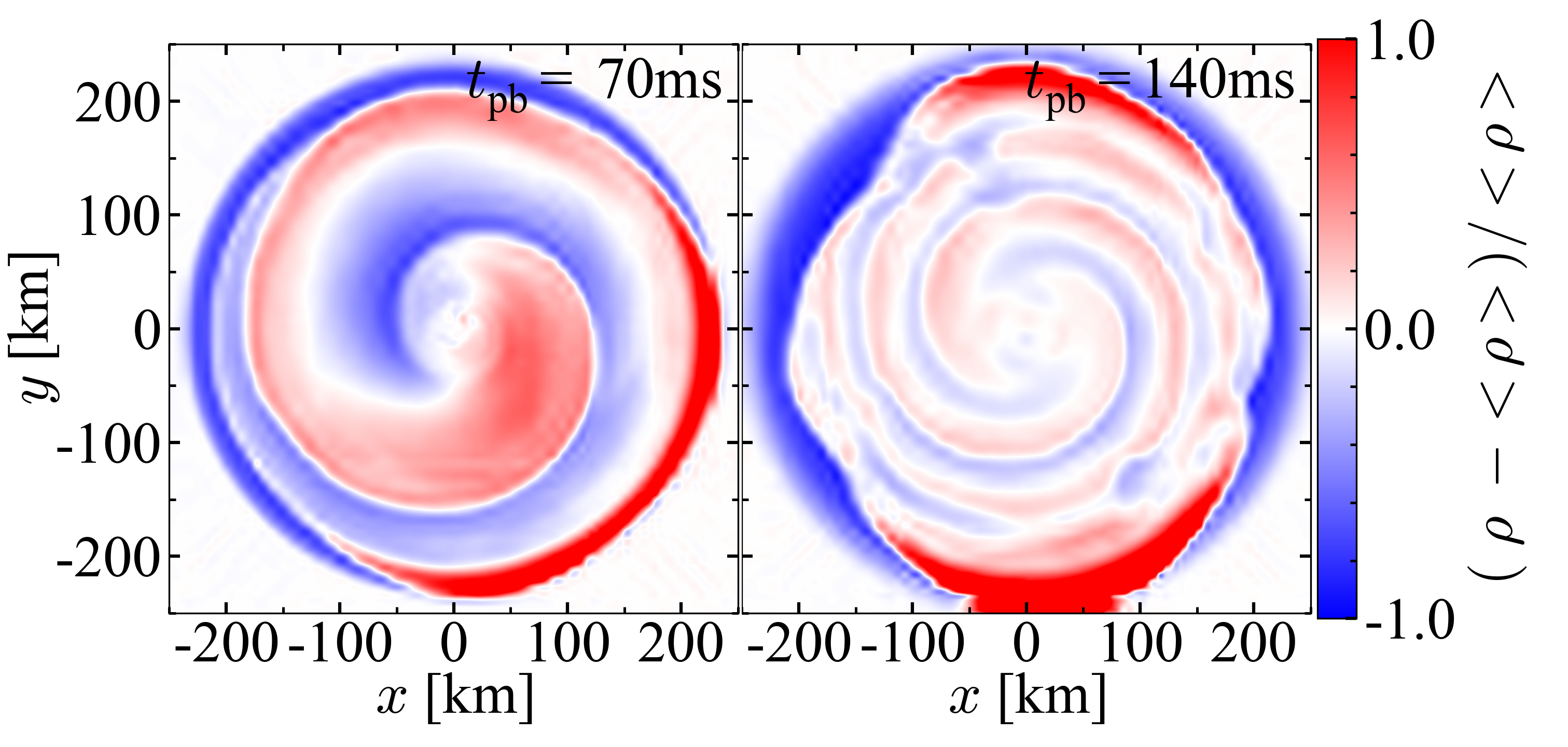}
\\
\includegraphics[bb= 720 10 2220 1600, width=0.235\textwidth]{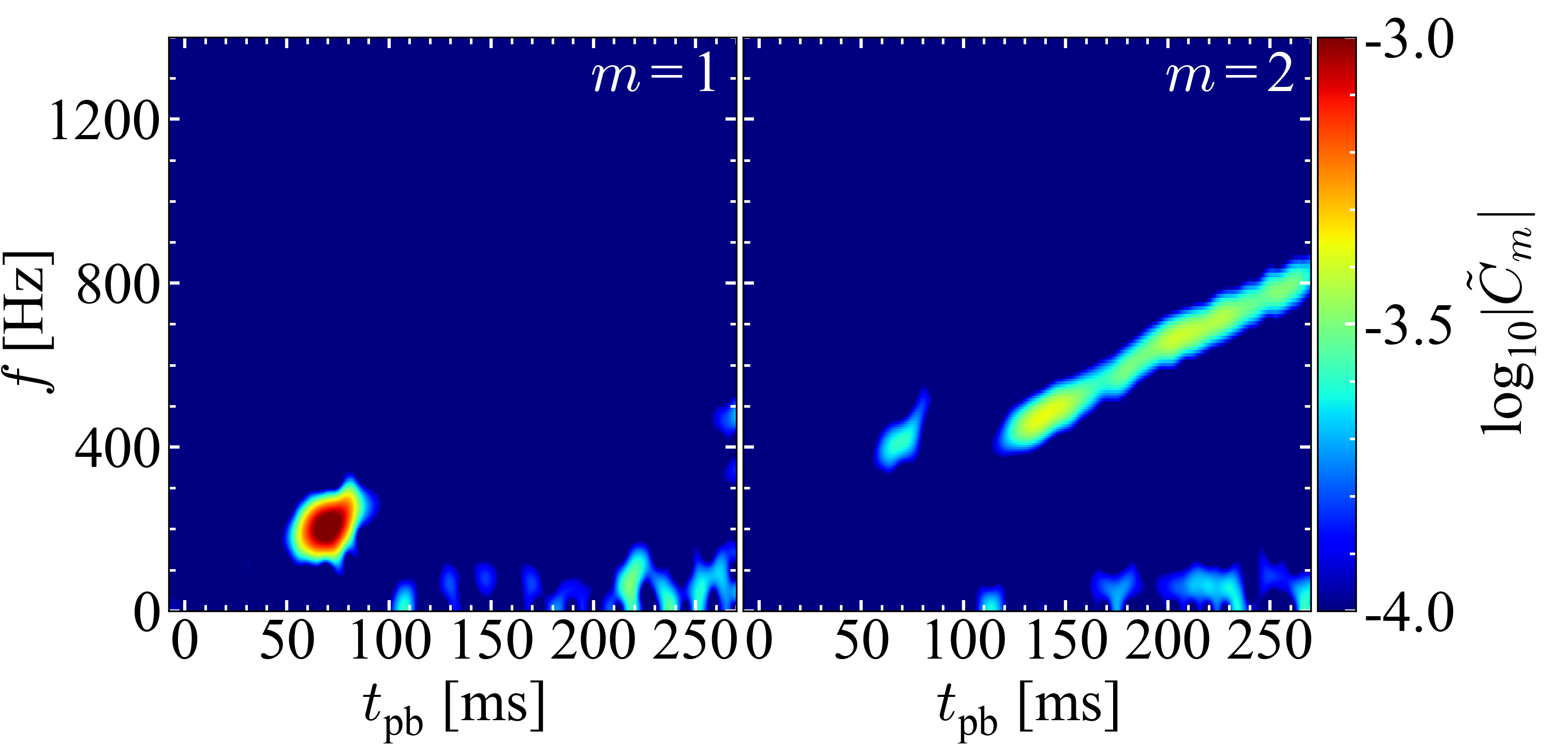}
\caption{Density deviation normalized by the averaged density in the equatorial plane $t_{\rm{pb}} = 70$ ms (top left) and $t_{\rm{pb}} = 140$ ms (top right) and spectrograms of $m=1$ (bottom left) and $m=2$ (bottom right) mode amplitudes for the density. The average is taken over a circular ring of radius $\varpi = \sqrt{x^2 + y^2}$ in the equatorial plane. The mode amplitude of $C_m$ is computed through the azimuthal Fourier decomposition of the density at $\varpi = 100$ km and $z = 0$ km and its spectrogram $\tilde{C}_{m}(t,f)$ is shown in the bottom panels for $m = 1, 2$. \label{fig:spiral}}
\end{figure}

Now we move on to investigate the emission mechanism of the quasi-periodic GW signal seen after $t_{\rm{pb}} \sim 60$ ms in our model.
In the top panels of Fig. \ref{fig:spiral}, we show the snapshots of the normalized density deviation from the angle averaged value, $(\rho - \left<\rho \right>)/\left<\rho \right>$, where $\left<\right>$ represents an angular average at a certain radius on the equatorial plane, at $t_{\rm{pb}} = 70$ ms ({\it left}) and $140$ ms ({\it right}).
 During the first phase of the quasi-periodic GW emission ($60 \lesssim t_{\rm{pb}} \lesssim 80$ ms), this one-armed flow is actually kept visible, whereas
 the two-armed spiral arms develop in the second active phase ($t_{\rm{pb}} \gtrsim 120$ ms) as seen in the top 
 panels of Fig. \ref{fig:spiral}. 
 In both of the phases, 
$T/|W|$ reaches $\sim 5$ \%.
 Indeed, this value is close to the onset condition of the low-$T/|W|$ instability as previously identified in the literature \citep{Ott05,Scheidegger10,takiwaki16}.

In order to clarify how the growth of the one- and two-armed spiral flows lead to the quasi-periodic GW emission, we perform an azimuthal Fourier decomposition of the density on the equatorial plane 
(at a radius of $\varpi\equiv (x^2+y^2)^{1/2} = 100$ km) 
as $C_{m}\left(t \right)=\int _0 ^{2\pi} \rho\, e^{i m \phi} d\phi / \int _0 ^{2\pi} \rho\,d\phi$, and obtain the spectrograms of $\tilde{C}_{m}(t,f)$ (using the real part).
In the bottom panels of Fig. \ref{fig:spiral}, we show $\tilde{C}_{m}(t,f)$ for $m = 1$ (left panel) and $m = 2$ mode 
(right panel), respectively. In the first active phase ($60\lesssim t_{\rm pb}\lesssim$80 ms), one can see that both the $m = 1$ and $2$ modes grow, but the $m = 1$ mode amplitude is bigger than the $m = 2$ mode. 
In the second active phase ($t_{\rm{pb}} \gtrsim 110$ ms),  the dominant mode is $m = 2$ as clearly seen from the bottom right panel of Fig. \ref{fig:spiral}.  Note that likewise the GW spectrogram (bottom right panel of Fig. \ref{fig:strain}),  the peak frequency of $\tilde{C}_{2}(t,f)$ (bottom right panel of Fig. \ref{fig:spiral}) also increases with time.

One can determine the eigenfrequency of the spiral-wave pattern (namely, the $m$-th mode of $f_{\rm{mode}, m}$)  from the peak frequency of $\tilde{C}_{m}(t, f)$.  Note that the frequency with respect to the pattern speed of the $m-$th mode is determined by $f_{\rm{pat}, m}=f_{\rm{mode}, m}/m$ \citep{Watts05}. So
$f_{\rm{mode}, 1}$ and $f_{\rm{pat}, 1}$ are identical for $m = 1$.
In the first active phase ($60\lesssim t_{\rm pb}\lesssim$80 ms), the $m = 1$ eigenfrequency is $f_{\rm{mode}, 1} = f_{\rm{pat}, 1} \sim 200\ $Hz (bottom left panel of Fig. \ref{fig:spiral}).
Similarly, in the second active phase, the $m = 2$ eigenfrequency at $f_{\rm{mode}, 2} \sim 400\ $Hz (bottom right panel of Fig. \ref{fig:spiral}) is, for instance,  translated into $f_{\rm{pat}, 2} \sim 200\ $Hz.

In both the first and second active phases,  it is a natural consequence that the $m=2$ mode frequency of $f_{\rm{mode}, 2}\,(= 2 
f_{\rm{pat}, 2})$ (i.e., bar-mode deformation of the spiral flows) leads to the dominant quadrupole GW emission with the same frequency. In fact, one can see a 
nice match of the ramp-up frequency feature  between the bar-mode amplitude ($\tilde{C}_2(t,f)$) of the spiral flows (bottom right panel of Fig. \ref{fig:spiral}) and the GW spectrogram (seen as a red band from $\sim 400$ to 800 Hz in the bottom right panel of Fig. \ref{fig:strain}).

\begin{figure}
\begin{center}
\includegraphics[bb= 0 25 470 360, width=0.47\textwidth]{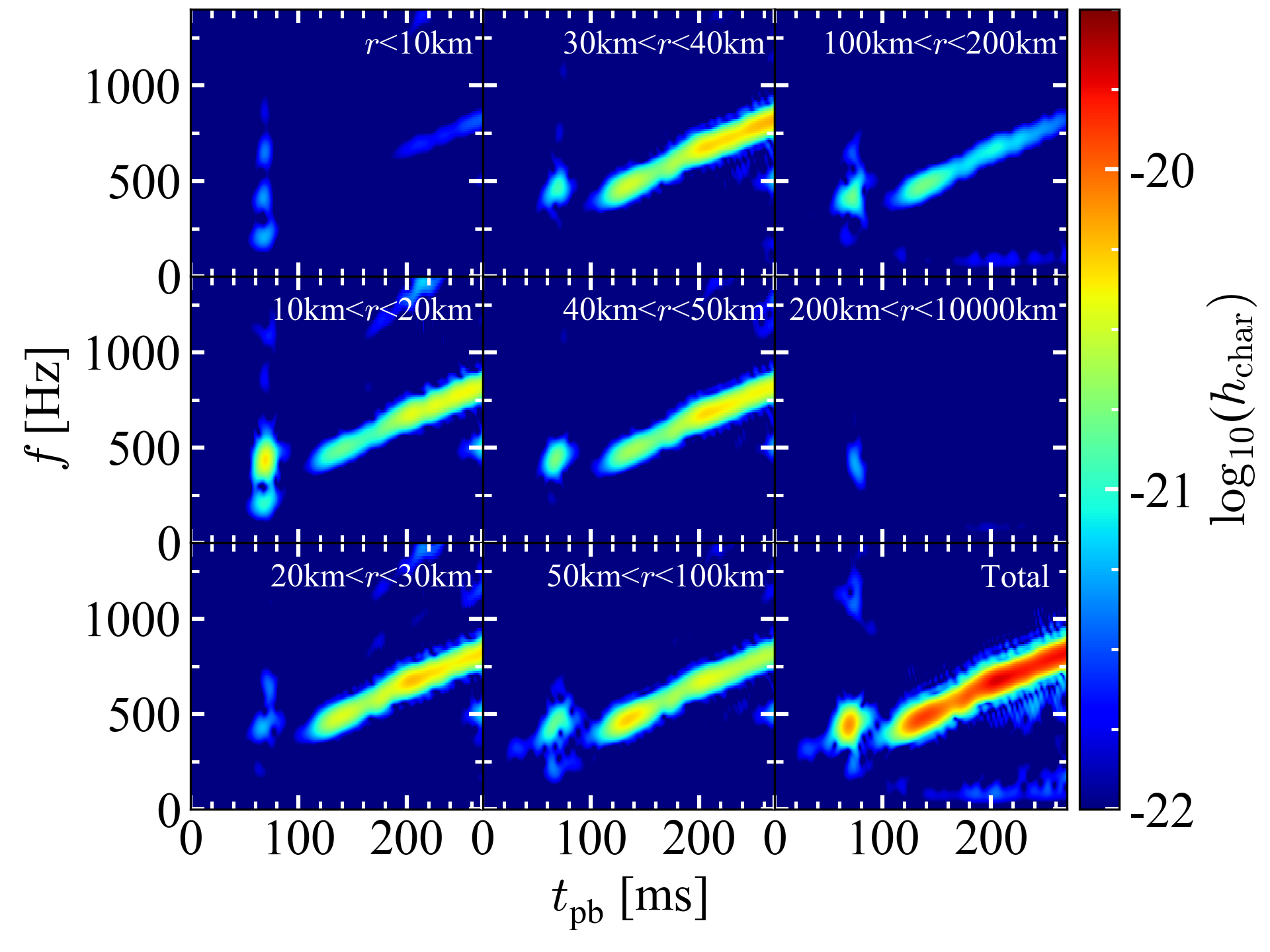}
\end{center}
\caption{Contributions from each spherical shell of radius $r$ to the GW spectrogram (seen from the pole) in a logarithmic scale of $h$. 
The range of the plotted layers is denoted in the upper right corner of each panel. 
As a reference, the bottom right panel shows the total GW spectrogram. \label{fig:layer}}
\end{figure}

Fig. \ref{fig:layer} shows contribution of different spherical shells to the GW spectrogram (seen from the pole). One can see that the ramp-up signature is generated almost all in the layers between $10 \lesssim r \lesssim 100$ km, whereas the dominant contribution from each shell differs with time.  Note that there is also a contribution from behind the shock ($100<r<200$ km).
Our results clearly show that both the non-axisymmetric flows that develop in the vicinity of the PNS core surface ($r\sim10$ km) and the spiral arms extending above coordinately give rise to the ramp-up GW emission.

\begin{figure}
\begin{center}
\includegraphics[bb= 0 20 470 300, width=0.47\textwidth]{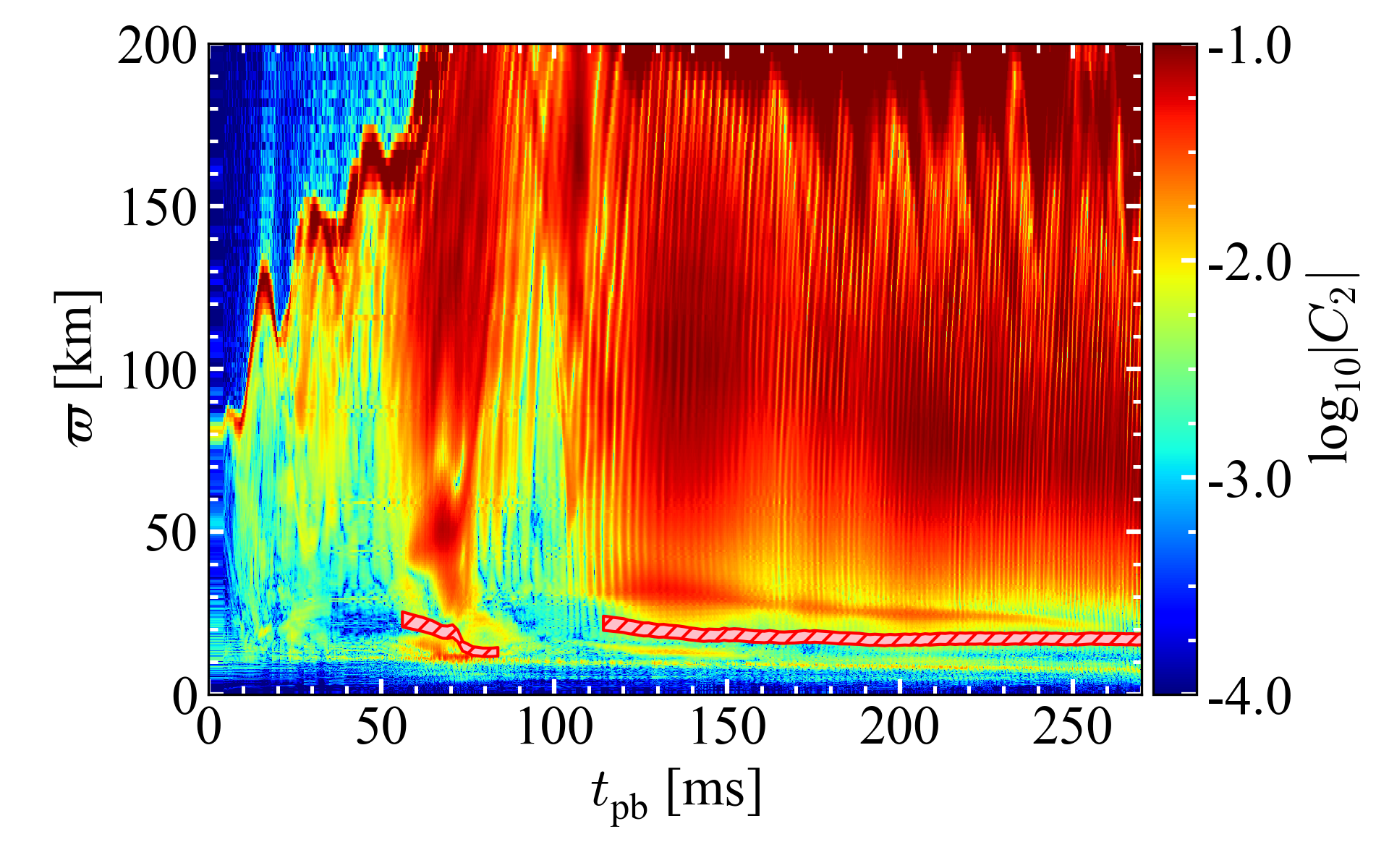}
\caption{Color map of $m=2$ mode amplitude for density variation $C_2$ as a function of time and cylindrical radius (see text for the definition). The red hatched band indicates $(1.0\pm 0.1) \varpi_{\rm{cor}}$, where $\varpi_{\rm{cor}}$ is the corotation radius.\label{fig:Cm}}
\end{center}
\end{figure}

 \citet{Watts05} firstly suggested that a necessary condition for the low-$T/|W|$ instability is the existence of the corotation radius where the angular velocity is equal to the pattern speed of an unstable mode \citep{saijo06}. Following 
 this idea, we attempt to interpret how the ramp-up feature seen in the second active phase ($t_{\rm pb}\gtrsim110$ ms) is produced.
 In order to clarify how the development of the $m = 2$ unstable mode is related to the corotation radius, we show in Fig. \ref{fig:Cm} the time evolution of spatial profile of normalized amplitude of the density perturbation with $m=2$ mode $|C_2(t,\varpi)|$ and the corotation radius ($\varpi_{\rm{cor}}$).  Note that there is a finite-width range ($\sim 10 \%$ level) in estimating the pattern speed
 (corresponding to the vertical width of the green
 stripe in the bottom right panel of Fig. \ref{fig:spiral}). Given this,  the location of the corotation radius is also defined with a 10 $\%$ error bar as $(1.0\pm 0.1)\, \varpi_{\rm{cor}}$, which is shown by the red hatched regions in Fig. \ref{fig:Cm}. As one can see from Fig. \ref{fig:Cm} and Fig. \ref{fig:layer}, the region above the corotation radius ($20$ km $\lesssim r\lesssim100$ km) contributes to the GW spectrogram more largely than the corotation radius ($10$ km $\lesssim r\lesssim20$ km) does. One of the reasons is that the highly deformed region (the reddish region in Fig. \ref{fig:Cm}), leading to stronger GW emission, locates above the corotation radius.
Seen from the hatched region of Fig.\ref{fig:Cm},  the corotation radius 
gradually shrinks from $\sim 22$ km ($t_{\rm pb} \sim 110$ ms)
 to $\sim 16$ km at the final simulation time. This closely coincides with the PNS core contraction.
As being dragged by the shrink of the corotation region, the inner edge of the highly deformed region with $m=2$ mode 
also moves inward.
This also supports the idea that the $m = 2$ PNS distortion may be generated via resonance  at the corotation radius, leading to the formation of the two-armed spiral waves.
The gradual recession of the corotation point leads to the spinning up the two-armed spiral waves (bottom right panel of Fig. \ref{fig:spiral}), which is reconciled with the ramp-up GW feature as seen in the bottom right panel of Fig. \ref{fig:strain}.

\begin{figure}
\begin{center}
\includegraphics[bb= 0 20 470 350, width=0.47\textwidth]{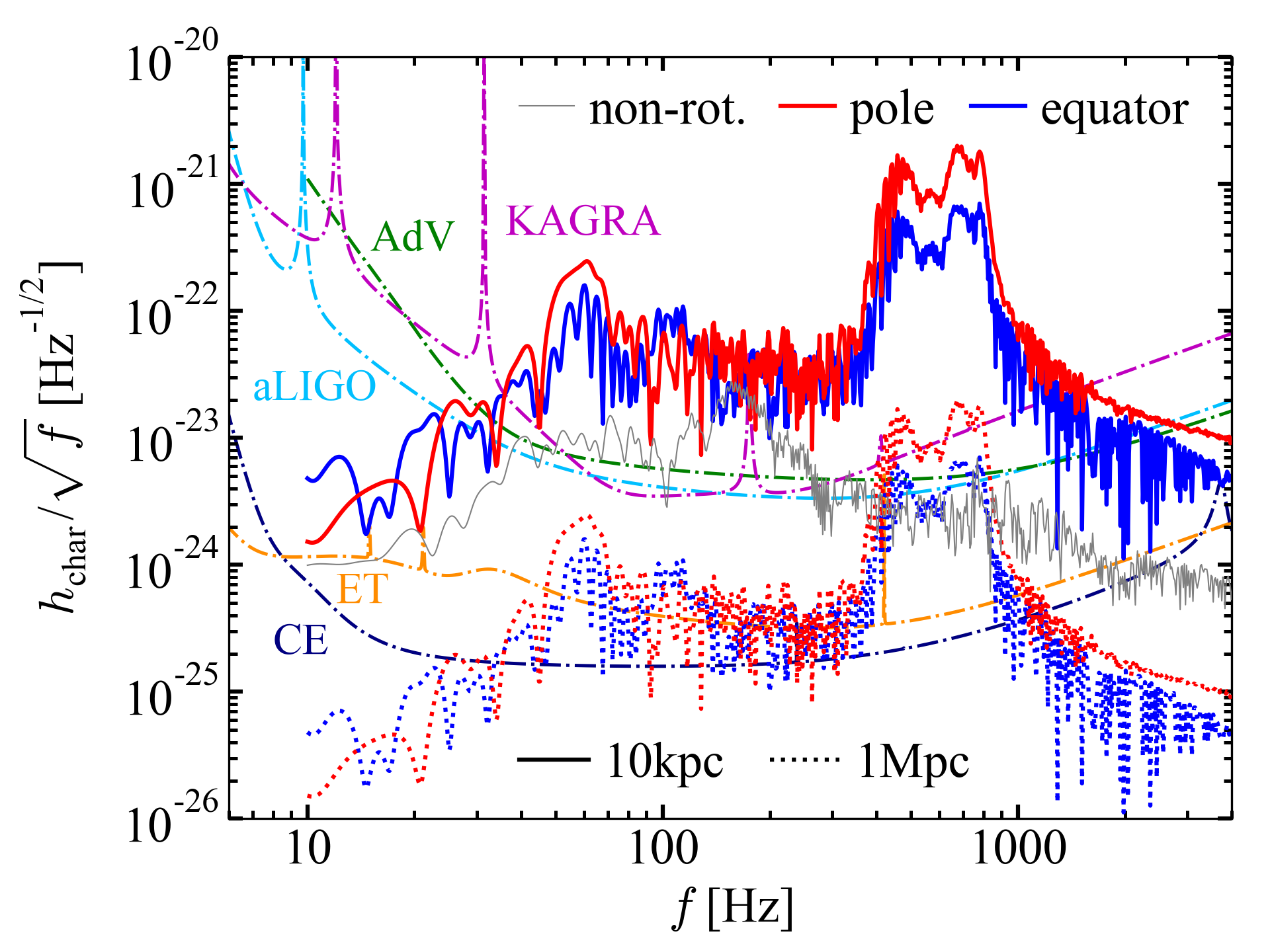}
\end{center}
\caption{Characteristic GW spectral amplitudes of our model seen along the pole (red lines) and along the equator (blue lines) and of the non-rotating model from \citet[][thin gray line]{KurodaT18} measured up to $t_{\rm{pb}}<270$ ms as a source distance of 10 kpc (solid lines) and 1 Mpc (dotted lines) relative to  the noise amplitudes of advanced LIGO (aLIGO; cyan), advanced VIRGO (AdV; green), KAGRA (magenta) from \citet{abbott18det}, Einstein Telescope \citep[ET; orange;][]{ET}, and Cosmic Explorer \citep[CE; navy;][]{CE}. The detector noise amplitudes are indicated by dash-dotted lines.}
\label{fig:det}
\end{figure}

In the end, we discuss detectability and observational rate of the GW signals.
Fig. \ref{fig:det} shows the GW spectral amplitudes seen from the polar (red lines) and the equatorial (blue lines) observer at  a distance of 10 kpc (solid lines) and 1 Mpc (dotted lines) relative to the sensitivity curves of the advanced LIGO, advanced VIRGO, and KAGRA \citep{abbott18det}; and the third-generation GW detectors of Einstein Telescope \citep{ET} and Cosmic Explorer \citep{CE}. The peak GW spectral amplitudes of our model are much larger than the non-rotating one of \citet[][thin gray line]{KurodaT18}. In accordance with the spectrogram analysis of Fig. \ref{fig:strain} and \ref{fig:spiral},
the peaks of the GW spectra are located around $400 \lesssim f \lesssim 900$ Hz, with the GW emission stronger toward the rotational axis (red line). 
 These GW signals can be a target of LIGO, Virgo, and KAGRA for a Galactic event. But more interestingly, these signals, the peak frequency of which is close to the best sensitivity range ($\lesssim$ 1kHz) of ET and CE, could be detectable out to Mpc distance scale by the third-generation detectors.
 
 To roughly estimate the observational rate of this kind of events, we make a bold assumption that a rapidly rotating massive star considered in this work would be associated with long gamma-ray bursts (lGRBs). About 1\% of massive stars would rotate at sufficient speeds for lGRB \citep{WH06,deMink13}. Among these stars, a fraction of $\lesssim$ 15\% would finally form BHs based on the assumption that all ultra metal-poor stars become BHs \citep{O'Connor11}. The CC rate in the Local Group (several Mpc) is estimated as 0.2 - 0.8 events yr$^{-1}$ \citep{nakamura16}. Therefore the rate for CC events like our model in the Local group would be estimated to be $\sim$ 0.0012 yr$^{-1}$.  Admitting that this event rate is an order of magnitude lower than the Galactic supernova event rate \citep[0.019$\pm$0.011 yr$^{-1}$ from][]{Diehl06}, our results demonstrate that detection of the {\it strongest} GW signals (so far predicted in the context of full GR neutrino-radiation hydrodynamics simulations) could provide a unique opportunity to probe into rapid rotation and the associated non-axisymmetric instabilities, otherwise obscured deep inside the massive stellar core.
\section{Discussions}
Finally we shall discuss several limitations in this work.
First our simulation does not take into account magnetic fields.
If magnetorotational instability (MRI) \citep[e.g.,][]{akiy03,Obergaulinger09,masada12,Rembiasz16a,Rembiasz16b} develops in a sufficiently short timescale, MRI could transfer angular momentum of the PNS outward \citep[e.g.,][]{Moesta15,masada15} and may prevent the growth of the non-axisymmetric instabilities. We have to tackle with this by developing full 3D GR-MHD code with spectral neutrino transport, however, this is beyond the scope of this work. Second, \citet{Saijo18} recently pointed out
 (in the context of the isolated NSs) that the use of different EOSs significantly impact the growth of the low-$T/|W|$ instability (e.g., \citet{Scheidegger10}). Not only the impact of EOS but also of updating the neutrino opacities \citep[e.g.,][]{bollig17} both of which could significantly affect the explodability and the PNS contraction remain to be investigated. 
Our full GR simulations can follow evolution in the vicinity of rapidly rotating PNSs more properly than simplified GR treatments like effective GR potentials \citep[e.g.][]{Marek06}, which is often constructed on the basis of a spherically symmetric space. We thus believe that our full GR treatment exerts its potential to derive quantitatively, or perhaps even qualitatively, correct GWs as they can be sensitive to the connection between the PNS core contraction and the low $T/|W|$ instability.
Finally a major limitation is that we have shown results of only one simulation sample.
Systematic study changing the progenitor mass, metallicity, rotation and magnetic fields is mandatory in order to draw a robust conclusion
 to clarify under which condition the strong GW emission found in this work can be obtained. In the decade to come, we speculate that these issues are going to be tackled by utilizing next-generation (Exa-scale) supercomputing resources.

\vspace{-0.5cm}
\section*{Acknowledgements}
This work has been partly supported by Grant-in-Aid for Scientific Research 
(JP17H01130, 
JP17K14306, 
JP18H01212) 
from the Japan Society for Promotion  of Science (JSPS) and the Ministry of Education, Science and Culture of Japan (MEXT, Nos. 
JP17H05206, 
JP17H06357, 
JP17H06364, 
JP24103001), and by the REISEP at Fukuoka University and the associated projects (Nos.\ 171042,177103),
and JICFuS as a priority issue to be tackled by using Post `K' Computer. 
TK acknowledges support from the European Research Council (ERC; FP7) under ERC Starting Grant EUROPIUM-677912.
Numerical computations were carried out on Cray XC30 and XC50 at Center for Computational Astrophysics, NAOJ, and on Cray XC40 at Yukawa Institute for Theoretical Physics, Kyoto University.

\bibliographystyle{mnras}
\bibliography{mybib.bib} 


\bsp    
\label{lastpage}
\end{document}